\documentclass[aps,pra,superscriptaddress,reprint]{revtex4-2}
\usepackage{amsmath}
\usepackage{amssymb}
\usepackage{mathrsfs}
\usepackage{graphicx}
\usepackage{booktabs}  
\usepackage{multirow}  
\usepackage{xcolor}    

\usepackage[colorlinks=true, letterpaper=true, pdfstartview=FitV, linkcolor=blue, citecolor=blue, urlcolor=blue]{hyperref}

\begin{document}
	
	\title{Nonreciprocal optomechanical entanglement in an asymmetric Fabry-P\'{e}rot cavity}
	
	\author{Jia-Kang Wu}
	\affiliation{Key Laboratory of Low-Dimensional Quantum Structures and
		Quantum Control of Ministry of Education, Key Laboratory for Matter Microstructure and Function of Hunan Province, Department of Physics and Synergetic Innovation Center for Quantum Effects and Applications, Hunan Normal University, Changsha 410081, China}
	
	\author{Ning Hu}
	\affiliation{Key Laboratory of Low-Dimensional Quantum Structures and
		Quantum Control of Ministry of Education, Key Laboratory for Matter Microstructure and Function of Hunan Province, Department of Physics and Synergetic Innovation Center for Quantum Effects and Applications, Hunan Normal University, Changsha 410081, China}
	
	\author{Jie-Qiao Liao}
	\email{jqliao@hunnu.edu.cn}
	\affiliation{Key Laboratory of Low-Dimensional Quantum Structures and
		Quantum Control of Ministry of Education, Key Laboratory for Matter Microstructure and Function of Hunan Province, Department of Physics and Synergetic Innovation Center for Quantum Effects and Applications, Hunan Normal University, Changsha 410081, China}
	\affiliation{Institute of Interdisciplinary Studies, Hunan Normal University, Changsha, 410081, China}
	\affiliation{Hunan Research Center of the Basic Discipline for Quantum Effects and Quantum Technologies, Hunan Normal University, Changsha 410081, China}
	
	\author{Xun-Wei Xu}
	\email{xwxu@hunnu.edu.cn}
	\affiliation{Key Laboratory of Low-Dimensional Quantum Structures and
		Quantum Control of Ministry of Education, Key Laboratory for Matter Microstructure and Function of Hunan Province, Department of Physics and Synergetic Innovation Center for Quantum Effects and Applications, Hunan Normal University, Changsha 410081, China}
	\affiliation{Institute of Interdisciplinary Studies, Hunan Normal University, Changsha, 410081, China}
	\affiliation{Hunan Research Center of the Basic Discipline for Quantum Effects and Quantum Technologies, Hunan Normal University, Changsha 410081, China}
	
	\date{\today}
	
	\begin{abstract}
		Nonreciprocal transmission (classical nonreciprocity) in  optomechanical systems based on asymmetric Fabry-P\'{e}rot (F-P) cavities has been theoretically proposed and experimentally demonstrated. However, nonreciprocal quantum effects, particularly nonreciprocal quantum entanglement, remain unexplored in such systems. Here, we propose to generate nonreciprocal optomechanical entanglement in an asymmetric F-P cavity and discuss the connection between the nonreciprocal transmission and nonreciprocal quantum entanglement. We reproduce the nonreciprocal transmission spectra by solving the quantum Langevin equations, and then discuss the optimal parameters to achieve nonreciprocal optomechanical entanglement in the system. We show that a greater and more robust optomechanical entanglement can be approached in the asymmetric F-P cavities, in comparing with the symmetric cavities. Furthermore, we find that the degrees of classical and quantum nonreciprocities do not exhibit positive correlation as expected. Our work shows that the classical and quantum nonreciprocities can be realized simultaneously in the asymmetric F-P cavities, which provide a platform to explore the connection between classical and quantum nonreciprocities.
	\end{abstract}
	
	\maketitle
	
	\section{Introduction}
	Nonreciprocal optical devices are indispensable components in both classical~\cite{T.Mizumoto2014,Z.L.Deck2018,QiuChengwei2024,YoungbloodNathan2025} and quantum information processing tasks~\cite{P.Cappellaro2017,F.Sciarrino2019,JingHui2025_2}. However, the integration of magneto-optical nonreciprocal devices on photonic chips remains an outstanding challenge. Alternative strategies for breaking the reciprocity without any magnetic field in chip-scale photonic devices have attracted increasing research interests~\cite{K.Harish2017,A.Andrea2019,YouLi2020}. Optomechanical systems~\cite{F.Marquardt2014}, involving the coupling of optical and mechanical modes through radiation pressure, provide a remarkable platform to achieve magnet-free nonreciprocity~\cite{DongChunhua2016,A.Andrea2017,Weig2022}. Various mechanisms have been introduced to realize optical nonreciprocity in the optomechanical systems, such as asymmetric Fabry-P\'{e}rot (F-P) cavity composed of two mirrors with different reflectivities~\cite{M.Lipson2009}, microring with unidirectional pumping~\cite{P.Rabl2012,E.Verhagen2016,LiBaijun2019,TangZhixiang2023}, synthetic magnetism~\cite{LiYong2015,F.Marquardt2015,A.A.Clerk2015,LiuYuxi2016,T.J.Kippenberg2017,A.Alu2017,O.Painter2017,ChenAixi2020,DongChunHua2021,LiuYongChun2021,ZhengShibiao2022}, and stimulated Brillouin scattering~\cite{GuoGuangcan2015,G.Bahl2015,XiaoMin2020,XuXunwei2023_2}. Many nonreciprocal optical devices have been theoretically proposed and experimentally demonstrated, including optical isolator and circulator~\cite{D.Jalas2013,XiaoMin2016,J.D.Teufel2017,B.Gaurav2017,XuXunwei2018,E.Verhagen2018,YangLiu2019,GaoFeng2019}, nonreciprocal router~\cite{WangXiaoguang2018,DongChunhua2023,XuXunwei2023}, directional amplifier~\cite{J.Aumentado2018,DongChunhua2018,LiYong2018,M.A.Sillanpaa2019,LongGuilu2022}, and nonreciprocal frequency converter~\cite{LiuYuxi2016,LiZhen2017}.
	
	Recently, the nonreciprocal effects have been extended from the classical to quantum regimes, and some quantum nonreciprocal effects have been predicted, such as nonreciprocal photon blockade~\cite{JingHui2018,JingHui2019,ChenAixi2020_2,YiXuexi2023,LiangErjun2023,ZhuAidong2023,ZhuAidong2024}, nonreciprocal quantum squeezing~\cite{YangG.J.2021,ZhangTiancai2023,ZhangS.2024,GuoQ.2024,BaiS.Y.2025,TangJ.2026}, and nonreciprocal quantum entanglement~\cite{JingHui2020,JingHui2022,JingHui2024,LvXinyou2024,YangHuan2024}. Nonreciprocal quantum entanglement was first proposed in spinning optomechanical resonator~\cite{JingHui2020}, where the time-reversal symmetry of the system is broken by the Sagnac effect. In comparison with the reciprocal quantum entanglement, nonreciprocal quantum entanglement exhibits significant advantages in the robustness against the backscattering losses, which offer novel possibilities for noise-tolerant quantum processors and backaction-immune quantum sensors~\cite{JingHui2022}. Besides the spinning optomechanical resonator, nonreciprocal quantum entanglement has also been proposed in other systems, including waveguide quantum electrodynamics systems~\cite{ChengMutian2024} and cavity magnonic systems~\cite{YangRongcan2020,RenYalong2022,YeLiu2023,ChenAixi2023,D.Camelia2023,YeLiu2024,ChenAixi2024,WangFei2024,ChenAixi2024_2,ZhuAidong2025,ChenAixi2025,ZhangShou2025}.
	
	The optomechanical system consisting of an asymmetric F-P type cavity provides an ideal platform for exploring diverse nonreciprocal effects. Nonreciprocal transmission has been theoretically proposed in the optomechanical system with asymmetrical F-P cavity~\cite{M.Lipson2009}, which has been experimentally demonstrated in an optomechanical system formed by a suspended meta-surface and a fixed mirror~\cite{ZhangJianfa2022}. In addition to the nonreciprocal transmission~\cite{YangYaping2014,NiuYueping2021,J.R.Lawall2023,XueJukui2023}, many other nonreciprocal effects have also been explored in the asymmetric F-P cavity,  such as nonreciprocal phonon lasing~\cite{XiongHao2023}, nonreciprocal quantum statistics~\cite{ZhuShiyao2017}, and nonreciprocal photon blockade~\cite{YangYaping2022,YuChangshui2023,BaiChenghua2025}. It is worth mentioning that the nonreciprocal photon blockade was realized experimentally in the asymmetric F-P cavities with cold atoms~\cite{ZhangTiancai2024}. However, nonreciprocal optomechanical entanglement in asymmetric F-P cavities remains unexplored.
	
	Here we propose to generate nonreciprocal optomechanical entanglement in an asymmetric F-P type optomechanical cavity formed by a fixed end mirror and a vibrating end mirror. We reproduce  the nonreciprocal transmission spectra of the system given in Ref.~\cite{ZhangJianfa2022} by solving the quantum Langevin equations (QLEs), which provide a method to investigate the quantum effects in the system. Based on the QLEs, we find the nonreciprocal optomechanical entanglement in the asymmetric F-P cavity, albeit very weak. To obtain a stronger optomechanical entanglement, we investigate the impacts of these parameters on the optomechanical entanglement. We find that the asymmetric cavities exhibit superior performance to the symmetric cavities in both the magnitude of entanglement and the robustness against thermal noise. In addition, we discuss the association between nonreciprocal transmission and nonreciprocal quantum entanglement, and find that the degrees of the classical and quantum nonreciprocities do not exhibit positive correlation as expected.
	
	The rest of this paper is organized as follows. In Sec.~\ref{Model}, we introduce the Hamiltonian of the asymmetric F-P cavity, and show the equations for the mean values of operators and the covariance matrix. In Sec.~\ref{NT}, we investigate the nonreciprocal transmission spectra based on the QLEs. In Sec.~\ref{NOE}, by using the logarithmic negativity to characterize the quantum entanglement, we discuss the optimal parameter conditions for optomechanical entanglement and explore the advantages of the asymmetric cavity over the symmetric case. Finally, we conclude this work in Sec.~\ref{Con}.
	
	\section{Physical Model and Equations of Motion}\label{Model}
\begin{figure}
		\includegraphics[width=0.30 \textwidth]{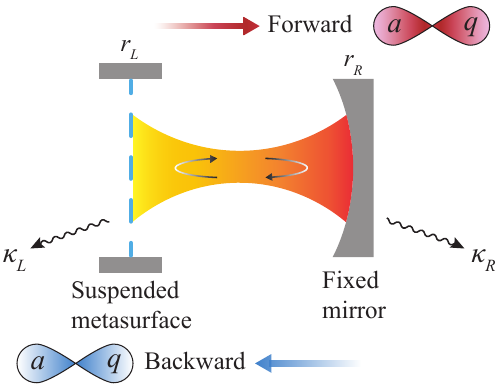}
		\caption{Schematic of the asymmetric F-P type optomechanical cavity, which is formed by a movable mirror (characterized by a reflectivity of $r_{L}$ and a decay rate of $\kappa_{L}$) and a fixed mirror (characterized by a reflectivity of $r_{R}$ and a decay rate of $\kappa_{R}$).}
		\label{Fig1}
	\end{figure}

	We consider an asymmetric F-P cavity consisting of a movable mirror formed by a suspended metasurface and a fixed mirror~\cite{ZhangJianfa2022}, as shown in Fig.~\ref{Fig1}. When the cavity is driven by a laser with frequency $\omega_{d}$, in a rotating frame with the unitary transformation operator $U=\exp(-i\omega_{d}a^{\dagger}at)$, the Hamiltonian of system can be written as
	\begin{eqnarray}
		\label{H}
		H &=& \hbar\Delta_{0} a^{\dagger}a + \frac{1}{2} \hbar\omega_{m} (p^2+q^2)+ \hbar\Omega (a^{\dagger} + a) \notag \\
		&&- \hbar g_{0} a^{\dagger}aq+ \frac{P_{L,\text{in}}+P_{L,\text{out}}}{c}x_{\text{zp}}q,
	\end{eqnarray}
	where $a$ ($a^{\dagger}$) is the annihilation (creation) operator of the cavity field mode (with resonance frequency $\omega_{c}$), $\Delta_{0}=\omega_{c}-\omega_{d}$ is the frequency
	detuning between the cavity mode and the pumping field, $q$ ($p$) is the dimensionless position (momentum) operator of the mechanical mode (with resonance frequency $\omega_{m}$). $\Omega=\sqrt{\kappa_{\sigma} P/\hbar\omega_{d}}$ is the driving amplitude, where $P$ is the incident laser power and $\kappa_{\sigma}$ is the decay rate. The fourth term in Eq.~(\ref{H}) represents the optomechanical coupling induced by the field in the cavity, where $g_{0}=(\omega_c/L_{\text{c}}) \sqrt{\hbar/m\omega_{m}}$ is the single-photon optomechanical coupling strength, $L_{\text{c}}$ is the cavity length and $m$ is the effective mass of the mechanical resonator. The last term in Eq.~(\ref{H}) represents the optomechanical interaction induced by the fields outside the movable mirror, where $P_{L,\text{in}}$ is the input power and $P_{L,\text{out}}$ is the output power, $c$ is the speed of light in the vacuum, $x_{\text{zp}}=\sqrt{\hbar/m\omega_{m}}$ is the zero-point fluctuation. The reflectivities of the suspended metasurface and the fixed mirror are denoted by $r_{L}$ and $r_{R}$, respectively. The reflectivity $r_{\sigma}$ of the mirror and the corresponding decay rate $\kappa_{\sigma}$ satisfy the relation (neglecting the cavity absorption)
	\begin{eqnarray}
		\kappa_{\sigma} = \frac{c}{2L_{\text{c}}} \frac{(1 - r_{\sigma})}{\sqrt{r_{\sigma}}}, \quad \sigma=L,R,
	\end{eqnarray}
	where the subscripts $L$ and $R$ denote the left (movable) mirror and right (fixed) mirrors, respectively.
	
	By considering both the dissipation and fluctuation of the operators, the QLEs of the system are given by
	\begin{align}
		\dot{q} =& ~\omega_{m} p, \\
		\dot{p} =& -\omega_{m}q -\frac{P_{L,\text{in}}+P_{L,\text{out}}}{c p_{\text{zp}}} + g_{0} a^{\dagger}a - \gamma_{m}p + \xi, \\
		\dot{a} =& -i\Delta_0 a + ig_{0}qa -\frac{\kappa_{\text{tot}}}{2} a -i\Omega \notag \\
		& + \sqrt{\kappa_{L}}a_{L,\text{in}} + \sqrt{\kappa_{R}}a_{R,\text{in}},
	\end{align}
	where $\gamma_{m}$ is the mechanical damping rate, $\kappa_{\text{tot}}=\kappa_{L}+\kappa_{R}$ is the total decay rate for the optical mode, and $p_{\text{zp}}=\sqrt{\hbar m\omega_{m}}$ is the standard deviation of zero-point momentum. The operators $a_{\sigma,\text{in}}$ and $\xi$ are the input noise operators associated with the optical mode and mechanical mode, respectively, which satisfy the following correlation functions
	\begin{align}
		\langle  a_{\sigma,\text{in}}(t) a^{\dagger}_{\sigma,\text{in}}(t') \rangle &= \delta(t-t'), \\
		\langle \xi(t) \xi(t') \rangle &\simeq \gamma_{m}(2\bar{n}+1)\delta(t-t'),
	\end{align}
	where $\bar{n}=[\exp(\hbar\omega_{m}/k_{B}T)-1]^{-1}$ denotes the mean thermal phonon number, $k_{B}$ is the Boltzmann constant, and $T$ is the temperature. Under the strong-driving condition ($|\Omega|\gg\kappa_{\text{tot}}$), the operators can be written as the sum of the mean values and quantum fluctuations, i.e., $a=\left\langle a \right\rangle_{ss}+\delta a$, $q=\left\langle q \right\rangle_{ss}+\delta q$, and $p=\left\langle p \right\rangle_{ss}+\delta p$. In the steady state, the mean values of operators are determined by the following equations
	\begin{align}
		\label{stv1}
		\left\langle p \right\rangle_{ss} &= 0, \\
		\label{stv2}
		\left\langle q \right\rangle_{ss} &= \frac{1}{\omega_{m}}\left[ g_{0}|\left\langle a \right\rangle_{ss}|^{2} -\frac{1}{p_{\text{zp}}c}(P_{L,\text{in}}+P_{L,\text{out}}) \right], \\
		\label{stv3}
		\left\langle a \right\rangle_{ss} &= \frac{i\Omega}{-i(\Delta_{0}-g_{0}\left\langle q \right\rangle_{ss})-\kappa_{\text{tot}}/2}.
	\end{align}
	
	By defining the quadrature fluctuations and input noises operators as $\delta X\equiv(\delta a+\delta a^{\dagger})/\sqrt{2}$, $\delta Y\equiv(\delta a-\delta a^{\dagger})/i\sqrt{2}$, $X_{\sigma,\text{in}} \equiv (a_{\sigma,\text{in}}+a_{\sigma,\text{in}}^{\dagger})/\sqrt{2}$, and $Y_{\sigma,\text{in}} \equiv (a_{\sigma,\text{in}}- a_{\sigma,\text{in}}^{\dagger})/i\sqrt{2}$, the QLEs for the quantum fluctuation operators can be written in a compact form as
	\begin{eqnarray}
		\dot{\mathbf{u}}(t)=\mathbf{A}\mathbf{u}(t)+\mathbf{v}(t),
	\end{eqnarray}
	where $\mathbf{u}(t) \equiv (q, p, X, Y)^{T}$ and $\mathbf{v}(t) \equiv (0, \xi, \sqrt{\kappa_{L}} X_{L,\text{in}} + \sqrt{\kappa_{R}} X_{R,\text{in}}, \sqrt{\kappa_{L}} Y_{L,\text{in}} + \sqrt{\kappa_{R}} Y_{R,\text{in}}$. The coefficient matrix $\mathbf{A}$ is given by
	\begin{eqnarray}
		\mathbf{A} = 
		\begin{pmatrix}
			0 & \omega_{m} & 0 & 0 \\
			-\omega_{m} & -\gamma_{m} & \text{Re}(G) & \text{Im}(G) \\
			-\text{Im}(G) & 0 & -\kappa_{\text{tot}}/2 & \Delta \\
			\text{Re}(G) & 0 & -\Delta & -\kappa_{\text{tot}}/2
		\end{pmatrix},
	\end{eqnarray}
	where $G=\sqrt{2}g_{0}\left\langle a \right\rangle_{ss}$ is the linearized optomechanical coupling rate, and $\Delta = \Delta_{0}-g_{0}^{2}|\left\langle a \right\rangle_{ss}|^{2}/\omega_{m}$ is the normalized detuning. When the system is stable, all eigenvalues of the matrix $\mathbf{A}$ must have negative real parts.
	
	The quantum correlations in the system can be characterized by the covariance matrix $\mathbf{V}$ with the elements $\mathbf{V}_{i,j}(t) \equiv [\left\langle \mathbf{u}_{i}(t)\mathbf{u}_{j}(t) \right\rangle +\left\langle \mathbf{u}_{j}(t)\mathbf{u}_{i}(t) \right\rangle]/2$ ($i,j=1,2,3,4$). Under the stability conditions, the solution for the covariance matrix can be expressed in the following form
	\begin{eqnarray}
		\label{Vij}
		\mathbf{V}_{i,j}=\sum_{k,l=1}^{4} \int_{0}^{\infty} ds \int_{0}^{\infty} ds' \mathbf{M}_{i,k}(s)\mathbf{M}_{j,l}(s')\mathbf{\Phi}_{k,l}(s-s'),\notag \\
	\end{eqnarray}
	where $\mathbf{M}(t)=\exp(\mathbf{A}t)$, and $\mathbf{\Phi}_{k,l}(s-s')=[\langle \mathbf{v}_{k}(s)\mathbf{v}_{l}(s') \rangle + \langle \mathbf{v}_{l}(s')\mathbf{v}_{k}(s) \rangle]/2$ is the stationary noise correlation functions. In the steady state ($t \rightarrow \infty$), the covariance matrix $\mathbf{V}$ fulfills the Lyapunov equation
	\begin{eqnarray}
		\label{Ly}
		\mathbf{AV}+\mathbf{VA}^{T} = -\mathbf{Q},
	\end{eqnarray}
	where $\mathbf{Q}=\text{Diag}[0,\gamma_{m}(2\bar{n}+1),\kappa_{\text{tot}}/2,\kappa_{\text{tot}}/2]$ is the diffusion matrix. The solution of $\mathbf{V}$ in the steady state can be obtained straightforwardly by solving Eq.~(\ref{Ly}).
	
	\section{Nonreciprocal transmission}\label{NT}
	
	\begin{figure}
		\includegraphics[width=0.48 \textwidth]{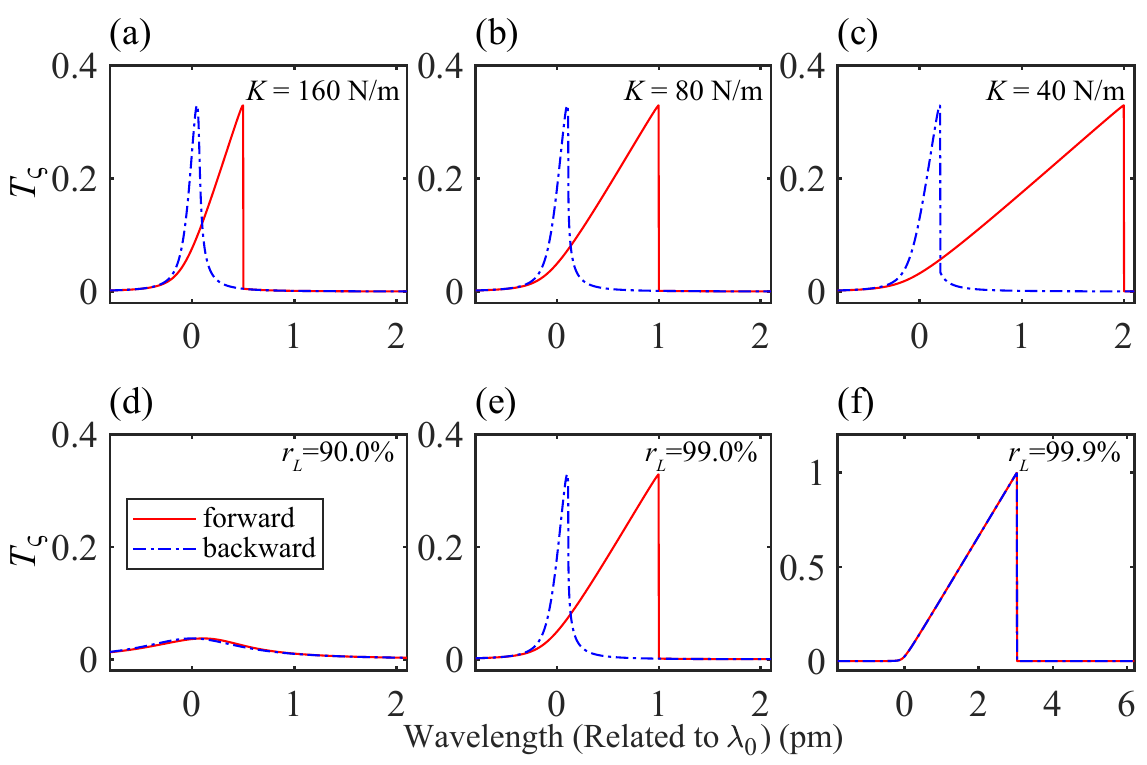}
		\caption{The effects of the spring constant $K$ and reflectivity $r_{L}$ on nonreciprocal transmission spectra. Red and blue curves represent the numerically calculated spectra for the forward and backward incidences, respectively. (a-c) Nonreciprocal transmission spectra of different spring constants $K$ with reflectivity fixed at $r_{L}=99.0\%$. (d-f) Nonreciprocal transmission spectra of different reflectivity $r_{L}$ with spring constant fixed at $K=80$ N/m. All other parameters remain identical in (a-f): the cavity length $L_{c}=17$ mm, the incident power $P=400$ mW, the incident wavelength is scanned upward from the shorter-wavelength region (left) with the central wavelength $\lambda_{0}=1550$ nm, and the mechanical resonance frequency $\omega_{m}/2\pi=114$ kHz.}
		\label{Fig2}
	\end{figure}
	
	In previous works on the optomechanical nonreciprocity based on the asymmetric F-P cavity~\cite{M.Lipson2009,ZhangJianfa2022}, the transmission spectra are simulated by the finite element method. However, this method cannot be employed to explore the quantum effects in the present system. The QLEs introduced in the last section provide as an uniform method to investigate both the transmission properties and the quantum effects.
	
	Based on the input-output relations~\cite{Collett1985}, the transmission rate of the system for the driving field input from the left mirror (forward driving) is given by
	\begin{align}
		\label{Tran1}
		T_{f} &= \frac{P_{R,\text{out}}}{P_{L,\text{in}}} = \frac{\hbar\omega_{d}\kappa_{R}|\left\langle a \right\rangle_{ss}|^{2}}{P_{L,\text{in}}}.
	\end{align}
	Similarly, the transmission rate for the driving field input from the right mirror (backward driving) is given by
	\begin{align}
		\label{Tran2} 
		T_{b} &= \frac{P_{L,\text{out}}}{P_{R,\text{in}}} = \frac{\hbar\omega_{d}\kappa_{L}|\left\langle a \right\rangle_{ss}|^{2}}{P_{R,\text{in}}}.
	\end{align}
	In order to quantify the magnitude of the classical nonreciprocity, we introduce the degree of classical nonreciprocity, which is defined by
	\begin{eqnarray}
		\eta_{c} = \frac{T_{f}-T_{b}}{T_{f}+T_{b}}.
	\end{eqnarray}
	When $\eta_{c}=0$, the system is classically reciprocal. If $\eta_{c}\neq 0$, then the system is classically nonreciprocal. In particular, $\eta_{c}=\pm 1$ correspond to the optimal classical nonreciprocity.
	
	In our numerical calculations, we use the same parameters in Ref.~\cite{ZhangJianfa2022}: the cavity length $L_{\text{c}}=17$ mm, the incident power $P=400$ mW, the mechanical resonance frequency $\omega_{m}/2\pi=114$ kHz, and the reflectivity of the right mirror fixed at $r_{R}=99.9\%$. The transmission spectra of the system are obtained by scanning the incident wavelength upward from the shorter-wavelength (left) to longer-wavelength regions (right) with the central wavelength $\lambda_{0}=1550$ nm, as shown in Fig.~\ref{Fig2}. The solid red curves and dashed blue curves correspond to the transmission rates for forward and backward drivings, respectively. As shown in Figs.~\ref{Fig2}(a)-\ref{Fig2}(c) for different values of the spring constant $K$, we find that the shift of the peaks between the forward and backward transmission increases with the decrease of the spring constant $K$ for a fixed reflectivity $r_{L}=99.0\%$. It means that the nonreciprocal behavior of transmission spectra becomes more obvious with the decrease of the spring constant $K$.
	
	For a fixed spring constant $K=80$ N/m, the transmission spectra with different $r_{L}$ are shown in Figs.~\ref{Fig2}(d)-\ref{Fig2}(f). When the reflectivity $r_{L}=90.0\%$, the transmission spectra for the forward and backward drivings are almost identical. This is because the low reflectivity of the metasurface will result in a weak optical field in the cavity. With the increase of the reflectivity $r_{L}$, a pronounced nonreciprocal behavior is observed at $r_{L}=99.0\%$. As the reflectivity $r_{L}$ further increases to $99.9\%$, the transmission spectra are completely coincident for both the forward and backward drivings. This is because the system becomes a symmetric cavity when $r_{L}=r_{R}$, leading to the disappearance of nonreciprocal behavior in the transmission spectra.
	
	The transmission spectra demonstrate an excellent agreement with those reported in Ref.~\cite{ZhangJianfa2022}, which confirm that the calculation of the transmission rates based on the QLEs is valid. In the light of the above discussions, blow we will explore the optomechanical entanglement in the system by calculating the logarithmic negativity $E_{\mathcal{N},\varsigma}$ based on the QLEs.
	
	\section{Nonreciprocal optomechanical entanglement}\label{NOE}
	\begin{figure}
		\includegraphics[width=0.45 \textwidth]{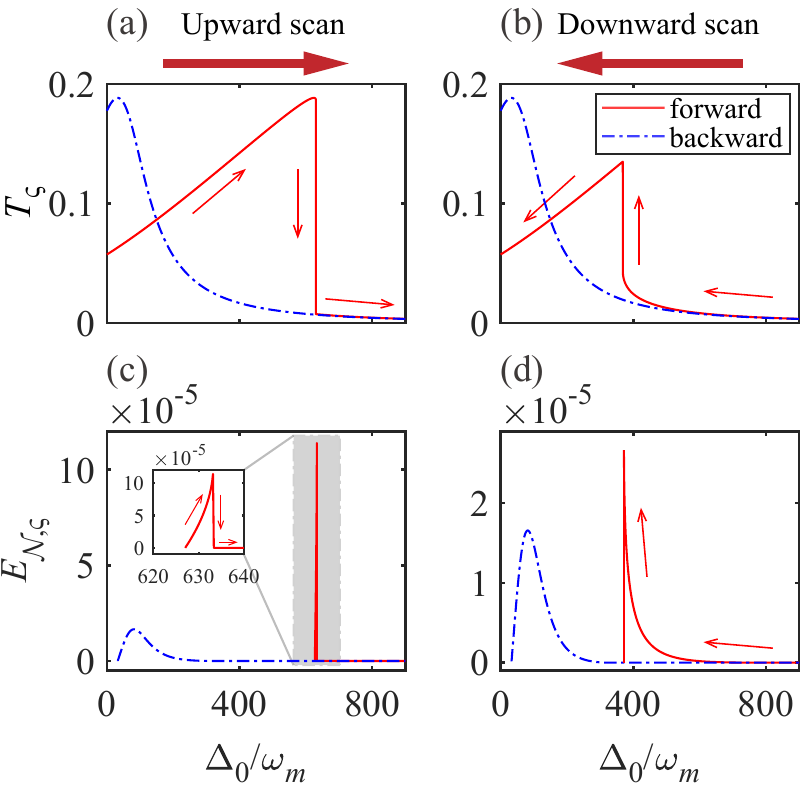}
		\caption{Nonreciprocal transmission spectra and logarithmic negativity $E_{\mathcal{N},\varsigma}$ versus the normalized detuning $\Delta_{0}/\omega_{m}$ under forward (red) and backward (blue) drivings for different scanning directions: (a) and (c) for upward frequency scanning; (b) and (d) for downward frequency scanning. The mechanical quality factor $\mathcal{Q}_{m}=3.1\times 10^{5}$, reflectivity $r_{L}=98.1\%$, spring constant $K=80$ N/m, and the temperature $T=10$ mK. Other parameters used are the same as Fig. \ref{Fig2}.}
		\label{Fig3}
	\end{figure}
	To characterize the entanglement between the optical and mechanical modes in the system, we employ the logarithmic negativity $E_{\mathcal{N},\varsigma}$, defined as~\cite{Fabrizio2004}
	\begin{eqnarray}
		\label{EN}
		E_{\mathcal{N},\varsigma} = \max[0,-\ln(2\eta^{-})], \quad \varsigma = f,b,
	\end{eqnarray}
	where $\eta^{-1} \equiv 2^{-1/2}\left\lbrace \Sigma(\mathbf{V}) -[\Sigma(\mathbf{V})^2 - 4\text{det}\mathbf{V}]^{1/2} \right\rbrace^{1/2}$, $\Sigma(\mathbf{V}) \equiv \text{det}\mathbf{D}+\text{det}\mathbf{B}-2\text{det}\mathbf{C}$. The covariance matrix $\mathbf{V}$ is written as the $2\times 2$ block form
	\begin{eqnarray}
		\mathbf{V} \equiv
		\begin{pmatrix}
			\mathbf{D} & \mathbf{C} \\
			\mathbf{C}^{T} & \mathbf{B}
		\end{pmatrix},
	\end{eqnarray} 
	where $\mathbf{B}$, $\mathbf{C}$ and $\mathbf{D}$ are $2\times 2$ matrices. The $E_{\mathcal{N},f}$ and $E_{\mathcal{N},b}$ are the logarithmic negativities under the forward and backward drivings, respectively. The Gaussian state of the system emerges entanglement if and only if $\eta^{-}<1/2$. To quantify the magnitude of the quantum nonreciprocity, we define the degree of quantum nonreciprocity
	\begin{eqnarray}
		\eta_{q} = \frac{E_{\mathcal{N},f}-E_{\mathcal{N},b}}{E_{\mathcal{N},f}+E_{\mathcal{N},b}}.
	\end{eqnarray}
	When $\eta_{q}=0$, the system is quantum reciprocal. If $\eta_{q}\neq 0$, then the system is quantum nonreciprocal, and $\eta_{q}=\pm 1$ correspond to the optimal quantum nonreciprocity.
	
	To explore the classical and quantum nonreciprocities simultaneously, we plot the transmission spectra $T_{\varsigma}$ and logarithmic negativity $E_{\mathcal{N},\varsigma}$ versus normalized the detuning $\Delta_{0}/\omega_{m}$ in Fig.~\ref{Fig3} for both forward (red) and backward (blue) drivings. It is found that there coexist nonreciprocal transmission and nonreciprocal optomechanical entanglement in the system. Different from the nonreciprocal transmission, however, the nonreciprocal optomechanical entanglement exists only within a very narrow frequency range, and the nonreciprocal optomechanical entanglement is very weak ($E_{\mathcal{N},\varsigma}$ on the order of $10^{-5}$). 
	
	As the system works in the bistable regime for the forward driving, we show the transmission spectra $T_{\varsigma}$ and logarithmic negativity $E_{\mathcal{N},\varsigma}$ under different frequency scanning directions in Fig.~\ref{Fig3}. The optomechanical entanglement appears in a narrow frequency range near the jump point in the bistable regime for both upward and downward frequency scanning directions. While the system is in the monostable regime for the backward driving, and the logarithmic negativity $E_{\mathcal{N},b}$ changes continuously independent of the frequency scanning directions.
	
	In order to obtain the optomechanical entanglement in a wider frequency range and to discuss the results conveniently, we will optimize the parameters to enhance the optomechanical entanglement and keep the system in the optical monostable regime. To get the optimal parameters, we discuss the impact of the following factors on the optomechanical entanglement: mechanical resonance frequency $\omega_{m}$, metasurface reflectively $r_{L}$, driving power $P$, and temperature $T$.
	
	\subsection{Mechanical frequency}
	\begin{figure}
		\includegraphics[width=0.49
		\textwidth]{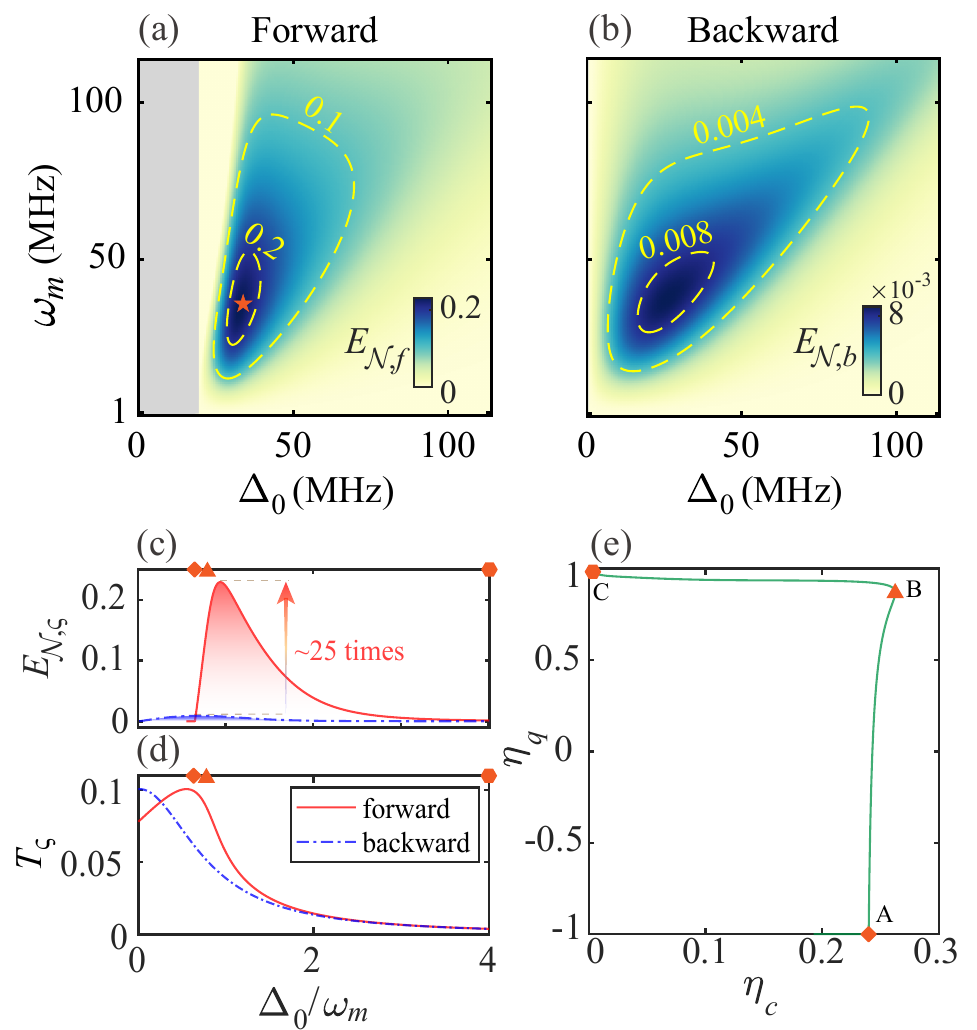}
		\caption{Logarithmic negativity $E_{\mathcal{N},\varsigma}$ versus $\Delta_{0}$ and $\omega_{m}$ under various drivings: (a) forward driving and (b) backward driving. The red star indicates the peak value of $E_{\mathcal{N},\varsigma}$ achieved under the forward driving, while the gray area denotes the unstable region. (c) Logarithmic negativity $E_{\mathcal{N},\varsigma}$ versus the scaled detuning $\Delta_{0}/\omega_{m}$ at the mechanical frequency $\omega_{m}/2\pi=34.5$ MHz. (d) Transmission spectrum versus the scaled detuning $\Delta_{0}/\omega_{m}$. (e) Degree of classical nonreciprocity $\eta_{c}$ versus quantum nonreciprocity $\eta_{q}$. Other parameters used are: incident power $P=200$ mW, reflectivities $r_{L}=96.3\%$ and $r_{R}=99.9\%$, and temperature $T=10$ mK.}
		\label{Fig4}
	\end{figure}
	
	As the optomechanical entanglement depends on the mechanical resonance frequency, we plot the variation of the logarithmic negativity $E_{\mathcal{N},\varsigma}$ with the detuning $\Delta_{0}$ and the resonant frequency $\omega_{m}$ in Figs.~\ref{Fig4}(a) and \ref{Fig4}(b). The gray area denotes the unstable region, and the red star marks the maximum of the logarithmic negativity $E_{\mathcal{N},f}$ where $\omega_{m}=34.5$ MHz. The maximal optomechanical entanglement $E_{\mathcal{N},f}$ reaches $\sim$0.2 and the maximal optomechanical entanglement $E_{\mathcal{N},b}$ only reaches $\sim$0.01, which means the optomechanical entanglement exhibits good nonreciprocity.
	
	Figure \ref{Fig4}(c) shows the logarithmic negativity $E_{\mathcal{N},\varsigma}$ versus the normalized detuning $\Delta_{0}/\omega_{m}$ for the mechanical frequency of $\omega_{m}=34.5$ MHz. The red and blue curves represent $E_{\mathcal{N},\varsigma}$ under the forward and backward drivings, respectively. The optomechanical entanglement universally exhibits an initial increase followed by a decrease as the normalized detuning $\Delta_{0}/\omega_{m}$ grows, and the maximal optomechanical entanglement located at $\Delta_{0}/\omega_{m}\sim 1$. It is found that the maximal optomechanical entanglement under the forward driving exceeds that under the backward driving by a factor of 25.
	
	To explore the association between the quantum nonreciprocity and classical nonreciprocity, we plot the variation of the transmission spectra with the normalized detuning $\Delta_{0}/\omega_{m}$ in Fig.~\ref{Fig4}(d). The red and blue curves represent the transmission rates $T_{\varsigma}$ under the forward driving and  backward driving, respectively. As the normalized detuning $\Delta_{0}/\omega_{m}$ increases, the maximum transmission rate of $T_{f}$ located at $\Delta_{0}/\omega_{m}\sim 0.6$, while $T_{b}$ continuously decreases.
	
	In order to show the association between the quantum nonreciprocity and classical nonreciprocity more intuitively, we plot the degree of quantum nonreciprocity $\eta_{q}$ and classical nonreciprocity $\eta_{c}$ in Fig.~\ref{Fig4}(e), based on the results in Figs.~\ref{Fig4}(c) and \ref{Fig4}(d). In the region from A to B, the degree of quantum nonreciprocity changes significantly, while the degree of classical nonreciprocity changes only slightly. In contrast, between the points B and C, there is a dramatic change in the degree of classical nonreciprocity $\eta_{c}$, whereas the degree of quantum nonreciprocity $\eta_{q}$ shows little change. It means that the degrees of classical and quantum nonreciprocities do not exhibit the positive correlation.
	
	\subsection{Reflectivity of the metasurface}
	\begin{figure}
		\includegraphics[width=0.45 \textwidth]{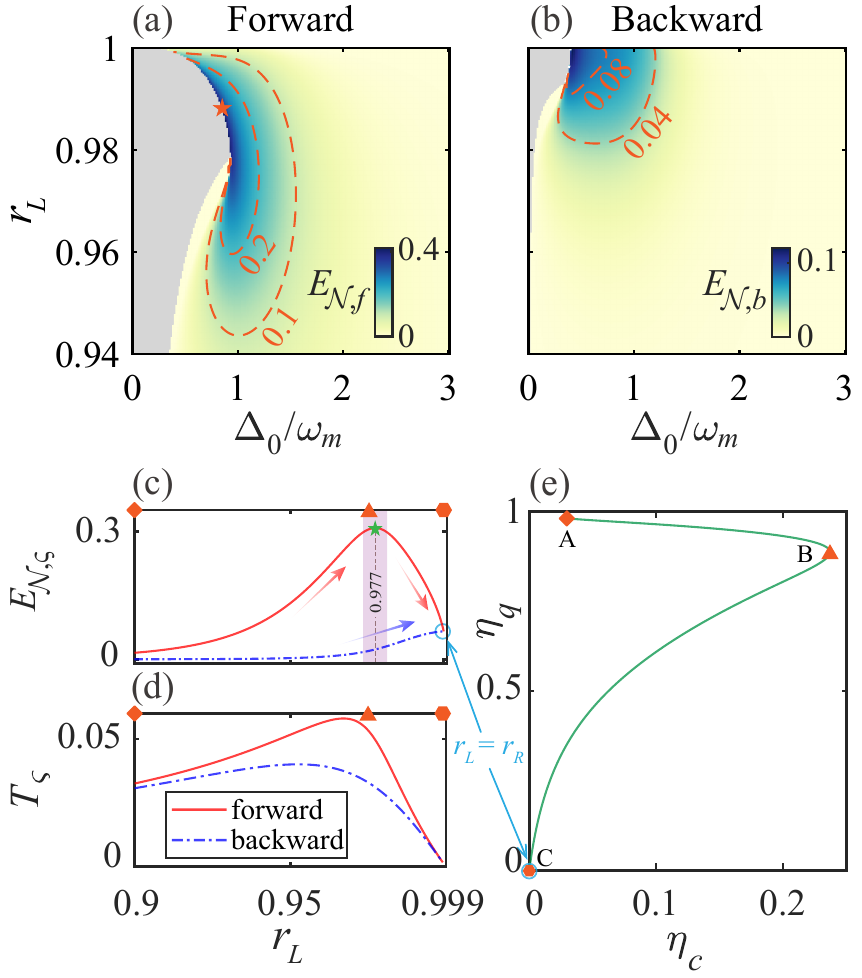}
		\caption{Logarithmic negativity $E_{\mathcal{N},\varsigma}$ versus $\Delta_{0}/\omega_{m}$ and reflectivity $r_{L}$ under either the (a) forward driving or the (b) backward driving. (c) Logarithmic negativity $E_{\mathcal{N},\varsigma}$ versus reflectivity $r_{L}$ under the forward and backward drivings. (d) Transmission spectra versus the reflectivity $r_{L}$ under the forward (red) and backward (blue) drivings with the normalized detuning $\Delta_{0}/\omega_{m}=1$. (e) Degree of classical nonreciprocity $\eta_{c}$ versus quantum nonreciprocity $\eta_{q}$. Other parameters used are the same as those in Fig.~\ref{Fig4}.}
		\label{Fig5}
	\end{figure}
	We now investigate the impact of the reflectivity $r_{L}$ on the optomechanical entanglement. Figures~\ref{Fig5}(a) and \ref{Fig5}(b) show the variation of the logarithmic negativity $E_{\mathcal{N},\varsigma}$ with the normalized detuning $\Delta_{0}/\omega_{m}$ and the reflectivity $r_{L}$ under the forward and backward drivings, respectively. For the backward driving, the maximal optomechanical entanglement ($E_{\mathcal{N},b} \sim0.1$) occurs at the perfect reflectivity $r_{L}=100\%$. The maximum optomechanical entanglement ($E_{\mathcal{N},f} \sim 0.4$) occurs at the reflectivity $r_{L}=98.7\%$ (the red star) for the forward driving. As the stability of the system is fragile at the reflectivity $r_{L}=98.7\%$ for the forward driving, we will choose the reflectivity $r_{L}=96.3\%$ to show the nonreciprocal entanglement in the following discussions.
	
	Figure~\ref{Fig5}(c) shows the logarithmic negativity $E_{\mathcal{N},\varsigma}$ as a function of the reflectivity $r_{L}$. The red and blue curves represent the logarithmic negativity $E_{\mathcal{N},\varsigma}$ under the forward and backward drivings, respectively. We find an interesting association between the optomechanical entanglement and the reflectivity $r_{L}$ for the forward driving: with the increase of $r_{L}$, $E_{\mathcal{N},f}$ first increases and then decreases with the maximum value at $r_{L}=97.7\%$. While under the backward driving, the optomechanical entanglement $E_{\mathcal{N},b}$ continuously strengthens with the increase of $r_{L}$. When the reflectivity $r_{L}$ equals to $r_{R}$, the optomechanical entanglement is reciprocal, i.e., $E_{\mathcal{N},f}=E_{\mathcal{N},b}$ (blue circle).
	
	To investigate the association between the quantum nonreciprocity and classical nonreciprocity, we plot the variation of the transmission spectra $T_{\varsigma}$ with reflectivity $r_{L}$ in Fig.~\ref{Fig5}(d). With the increase of the reflectivity $r_{L}$, both the transmission rates $T_{f}$ and $T_{b}$ are first increases and then decreases, and we have $T_{f}>T_{b}$ in the region $0.9<r_{L}<0.999$.
	
	Based on the results in Figs.~\ref{Fig5}(c) and \ref{Fig5}(d), the degree of quantum nonreciprocity $\eta_{q}$ and classical nonreciprocity $\eta_{c}$ are shown in Fig. \ref{Fig5}(e). In the region from A to B, the degree of quantum nonreciprocity $\eta_{q}$ gradually decreases, whereas the degree of classical nonreciprocity $\eta_{c}$ rapidly increases. In the region between B and C, both the degrees of classical and quantum nonreciprocities gradually decrease.
	
	\subsection{Incident power}
	\begin{figure}
		\includegraphics[width=0.45 \textwidth]{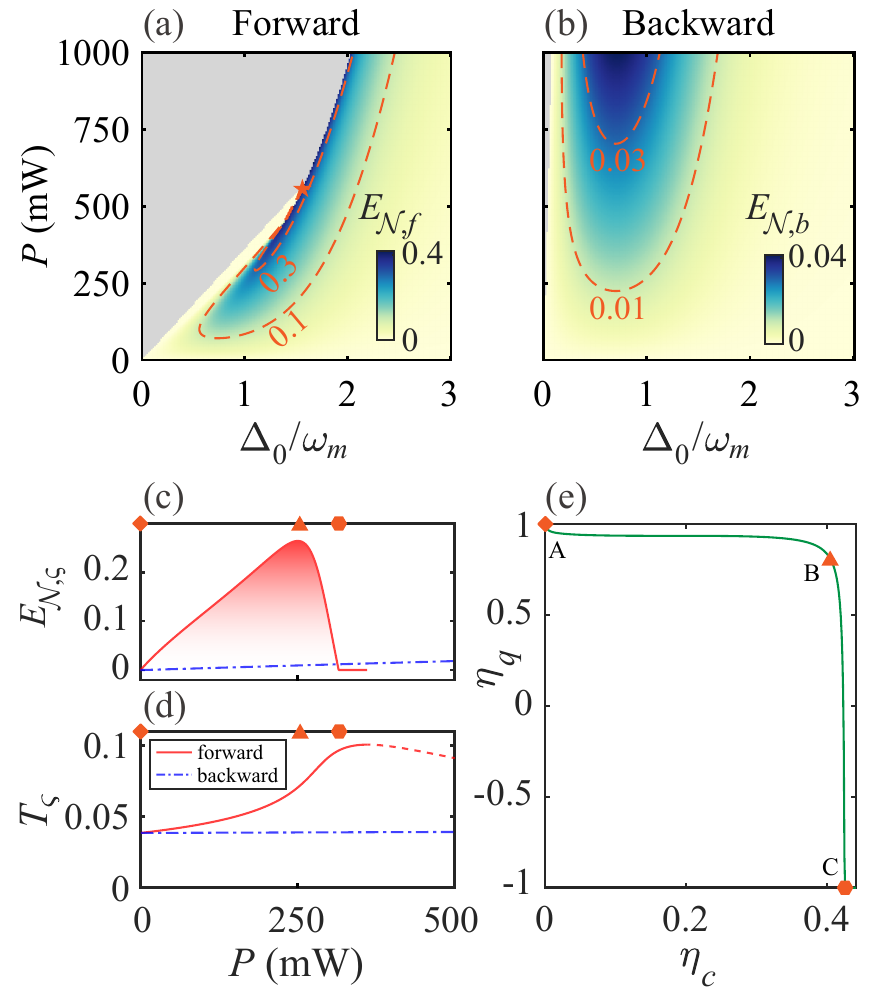}
		\caption{Logarithmic negativity $E_{\mathcal{N},\varsigma}$ versus $\Delta_{0}/\omega_{m}$ and $P$ under different drivings: (a) forward driving, (b) backward driving. (c) Logarithmic negativity $E_{\mathcal{N},\varsigma}$ versus incident power $P$ under the forward (red) and backward (blue) drivings. (d) Transmission spectra versus the incident power $P$ under the forward (red) and backward (blue) drivings. (e) Degree of classical nonreciprocity $\eta_{c}$ versus quantum nonreciprocity $\eta_{q}$. The reflectivity of the metasurface is $r_{L}=96.3\%$. Other parameters used are the same as those in Fig.~\ref{Fig5}.}
		\label{Fig6}
	\end{figure}
	
	\begin{figure*}
		\includegraphics[width=0.83 \textwidth]{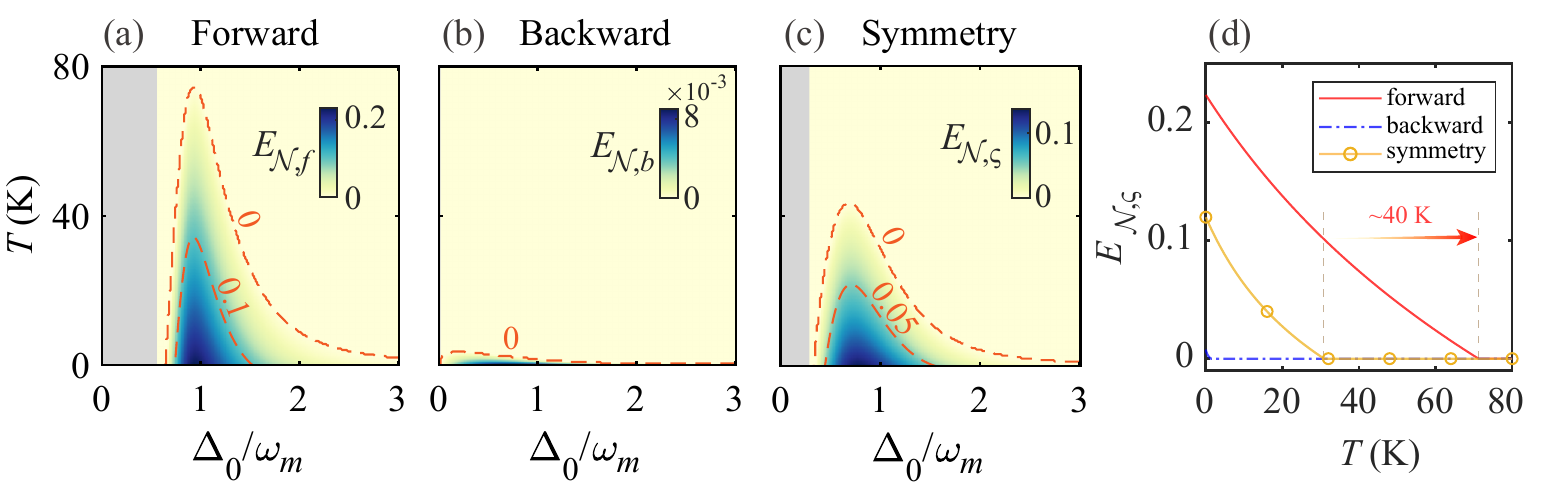}
		\caption{Logarithmic negativity $E_{\mathcal{N},\varsigma}$ versus  $\Delta_{0}/\omega_{m}$ and $T$ in both symmetric and asymmetric cavities under various driving cases: (a) forward driving, (b) backward driving, and (c) symmetric cavity. (d) Logarithmic negativity $E_{\mathcal{N},\varsigma}$ versus the temperature $T$ in both symmetric and asymmetric cavities. The red and blue curves represent $E_{\mathcal{N},\varsigma}$ under the forward and backward drivings in the asymmetric cavity, respectively, while the yellow curve represents $E_{\mathcal{N},\varsigma}$ in the symmetric cavity under the same driving conditions. The incident power is $P=200$ mW. The reflectivity in different cases is set as follows: for the asymmetric cavity, $r_{L}=96.3\%$ and $r_{R}=99.9\%$; for the symmetric cavity, $r_{L}=r_{R}=98.1\%$. Other parameters used are the same as those in Fig.~\ref{Fig6}.}
		\label{Fig7}
	\end{figure*}
	
	The effective optomechanical coupling $G$ linearly depends on $\left\langle a \right\rangle_{ss}$, which is determined by the incident power $P$. Figures \ref{Fig6}(a) and \ref{Fig6}(b) show the variation of the logarithmic negativity $E_{\mathcal{N},\varsigma}$ with the normalized detuning $\Delta_{0}/\omega_{m}$ and the incident power $P$. It is found that the maximal value of $E_{\mathcal{N},f}$ is $0.42$, which is about 10 times that of $E_{\mathcal{N},b}$ ($\sim 0.04$). With the increase of the incident power $P$, the optomechanical entanglement for the forward driving reaches its maximum at $P=558$ mW, as shown by the red star in Fig.~\ref{Fig6}(a). But the system is close to the unstable region under this driving power. Therefore, to maintain a strong optomechanical entanglement while ensuring the stability of system, we select the driving power $P=200$ mW in the following discussions.
	
	As shown in Figs.~\ref{Fig6}(c) and \ref{Fig6}(d), both the logarithmic negativity $E_{\mathcal{N},\varsigma}$ and the transmission spectra $T_{\varsigma}$ are plotted as functions of the incident power $P$ with a fixed normalized detuning $\Delta/\omega_{m}=1$. For the forward driving, the maximal value of the logarithmic negativity $E_{\mathcal{N},f}=0.26$ is achieved with the incident power $P=250$ mW, and the system becomes unstable when $P>360$ mW. For the backward driving, the system is stable for $0<P<500$ mW, and the logarithmic negativity $E_{\mathcal{N},b}$ rises slowly with the increasing incident power $P$, with the maximal optomechanical entanglement $E_{\mathcal{N},b}=0.02$ for $P=500$ mW. Different from the optomechanical entanglement, the maximal transmission rate $T_{f}=0.1$ is located at $P\sim 360$ mW, whereas $T_{b}$ remains nearly constant for $0<P<500$ mW.
	
	The association between the degree of quantum nonreciprocity $\eta_{q}$ and classical nonreciprocity $\eta_{c}$ is shown in Fig.~\ref{Fig6}(e) by using the data in Figs.~\ref{Fig6}(c) and \ref{Fig6}(d). It can be seen that, between the points A to B, $\eta_{q}$ changes slowly while $\eta_{c}$ increases significantly. In contrast, from B to C, $\eta_{q}$ decreases rapidly and $\eta_{c}$ shows little change.
	
	\subsection{Temperature}
	In this subsection, we discuss the robustness of optomechanical entanglement against temperature. Figures~\ref{Fig7}(a) and \ref{Fig7}(b) show the variation of the logarithmic negativity $E_{\mathcal{N},\varsigma}$ for the asymmetric cavity with the normalized detuning $\Delta_{0}/\omega_{m}$ and temperature $T$ under the forward and backward drivings, respectively. It is found that the maximal value of the logarithmic negativity $E_{\mathcal{N},f}$ near $\sim 0.2$, and the optomechanical entanglement for the forward driving persists to $T=74$ K with $\Delta_{0}/\omega_{m}=0.94$. For the backward driving, the maximal value of the logarithmic negativity $E_{\mathcal{N},b}$ only reach $\sim 0.01$, and the optomechanical entanglement persists to $T=3$ K with $\Delta_{0}/\omega_{m}=0.2$.
	
	To explore the advantages of the asymmetric cavity compared to the symmetric cavity, we show the logarithmic negativity $E_{\mathcal{N},\varsigma}$ versus the scaled detuning $\Delta_{0}/\omega_{m}$ and temperature $T$ for the symmetric cavity in Fig.~\ref{Fig7}(c). In order to ensure comparability, the total dissipation rate $\kappa_{\text{tot}}$ of the symmetric cavity equals to the asymmetric case. For the symmetric cavity, the maximal value of $E_{\mathcal{N},\varsigma}\sim 0.14$, and the optomechanical entanglement persists to $T=43$ K with $\Delta_{0}/\omega_{m}=0.7$.
	
	In order to show the advantages of the temperature robustness in the asymmetric cavity more intuitively, we plot the logarithmic negativity $E_{\mathcal{N},\varsigma}$ as a function of the temperature $T$ for both asymmetric and symmetric cavities with a fixed normalized detuning $\Delta_{0}/\omega_{m}=1$ in Fig.~\ref{Fig7}(d). The red and blue curves represent $E_{\mathcal{N},\varsigma}$ in the asymmetric cavity for the forward and backward drivings, respectively. The yellow curve denotes the logarithmic negativity $E_{\mathcal{N},\varsigma}$ in the symmetric cavity. We find that the optomechanical entanglement under the forward driving persists up to 71 K in the asymmetric cavity, compared to only 31 K in the symmetric cavity. It means that the optomechanical entanglement under the forward driving in the asymmetric cavity exhibits much stronger temperature robustness in comparison with the symmetric case. 
	
	\section{Conclusion}\label{Con}
	In conclusion, we have shown how to realize nonreciprocal optomechanical entanglement in an asymmetric F-P type optomechanical cavity. We have found the optimal parameters for the optomechanical entanglement, which enable both nonreciprocal optomechanical entanglement and nonreciprocal transmission to operate in the same parameter regime. We have also found that the correlation between the nonreciprocal transmission and nonreciprocal optomechanical entanglement is not always positive. Furthermore, our results demonstrate that the asymmetric optomechanical cavity offers a distinct advantage over the symmetric case, in achieving a greater and more robust optomechanical entanglement. With these advantages, the asymmetric cavity offer potential applications in quantum communication and quantum sensors with immunity against thermal noise. In addition, the asymmetric F-P cavity establish a new avenue towards exploring the connection between the quantum and classical nonreciprocities.
	
	\section*{Acknowledgments}
	This work was supported by the Innovation Program for Quantum Science and Technology (Grant No.~2024ZD0301000), the National Natural Science Foundation of China (Grants No.~12247105, No.~12421005 No.~12575015, and No.~12175061), the Sci-Tech Innovation Program of Hunan Province (Grant No.~2022RC1203), the Hunan Provincial Major Sci-Tech Program (Grant No.~2023ZJ1010), and the National Key Research and Development Program of China (Grant No. 2024YFE0102400).
	
	\bibliography{ref}
	
\end{document}